%% file: bubbles.tex
\documentclass[twocolumn]{emulateapj}
\usepackage{multirow}
\usepackage{natbib}
\begin{document}

\title{H {\scriptsize II} Region Driven Galactic Bubbles and Their Relationship to the Galactic Magnetic Field}
\author{Michael D. Pavel\altaffilmark{1} and D. P. Clemens}
\altaffiltext{1}{Current address: Department of Astronomy, The University of Texas at Austin, 2515 Speedway, C1400, Austin, TX 78712-1205}
\affil{Institute for Astrophysical Research}
\affil{Boston University, 725 Commonwealth Ave, Boston, MA 02215}
\email{email: pavelmi@bu.edu; clemens@bu.edu}

\slugcomment{Accepted for Publication in The Astrophysical Journal}

\shorttitle{Magnetic Alignment of Galactic Bubbles}
\shortauthors{Pavel, M. D., \& Clemens, D. P.}

\begin{abstract}
The relative alignments of mid-infrared traced Galactic bubbles are compared to the orientation of the mean Galactic magnetic field in the disk. The orientations of bubbles in the northern Galactic plane were measured and are consistent with random orientations - no preferential alignment with respect to the Galactic disk was found. A subsample of {\rm H\,{\scriptsize II}} region driven Galactic bubbles was identified, and as a single population they show random orientations. When this subsample was further divided into subthermal and suprathermal {\rm H\,{\scriptsize II}} regions, based on hydrogren radio recombination linewidths, the subthermal {\rm H\,{\scriptsize II}} regions showed a marginal deviation from random orientations, but the suprathermal {\rm H\,{\scriptsize II}} regions showed significant alignment with the Galactic plane. The mean orientation of the Galactic disk magnetic field was characterized using new near-infrared starlight polarimetry and the suprathermal {\rm H\,{\scriptsize II}} regions were found to preferentially align with the disk magnetic field. If suprathermal linewidths are associated with younger {\rm H\,{\scriptsize II}} regions, then the evolution of young {\rm H\,{\scriptsize II}} regions is significantly affected by the Galactic magnetic field. As {\rm H\,{\scriptsize II}} regions age, they cease to be strongly linked to the Galactic magnetic field, as surrounding density variations come to dominate their morphological evolution. From the new observations, the ratios of magnetic-to-ram pressures in the expanding ionization fronts were estimated for younger {\rm H\,{\scriptsize II}} regions.
\end{abstract}


\section{Introduction}
\label{4:intro}

Mid-infrared (MIR) objects called ``Galactic Bubbles'' were cataloged by \citet{2006ApJ...649..759C} and the sky projections of these three-dimensional bubbles were shown to be preferentially elliptical. What forces cause these bubbles to show non-circular shapes? External magnetic fields are one possible explanation, since ordered magnetic fields can apply anisotropic pressure. To test this hypothesis, the orientations of a subset of {\rm H\,{\scriptsize II}} region driven Galactic bubbles were compared to predictions of the mean Galactic magnetic field orientation.

Anisotropic bubbles can have a number of causes: stellar motions, Galactic shear, anisotropic driving forces (e.g., bipolar outflows), expansion into a non-uniform medium, or the presence of a magnetic field, which adds anisotropic pressure \citep[see][for a thorough review]{RevModPhys.67.661}. Detailed models for the expansion of stellar wind-driven bubbles into a uniform interstellar medium (ISM) were first developed by \citet{1975ApJ...200L.107C} and \citet{1977ApJ...218..377W}. The effects of magnetic fields on bubble expansion, in particular how the thickness changes at different points on the bubble's exterior shell, were considered by \citet{1991ApJ...375..239F} for supernovae and \citet{RevModPhys.67.661} for wind-driven bubbles. \citet{1992PASJ...44..177T} quantified the effect of a uniform magnetic field on a supernova remnant, predicting that the remnant should elongate along the magnetic field direction. This prediction has been confirmed for supernovae \citep[e.g.,][]{1998ApJ...493..781G}, but not for wind-driven or radiation-driven bubbles.

\citet{1977ApJ...218..377W} showed that neither stellar motions nor Galactic shear can account for the elongation of Galactic bubbles. The lifetimes of early-type stars responsible for creating these bubbles are much shorter than the timescales needed for stellar motion or Galactic shear to significantly warp the bubbles. Large stellar motions (e.g., runaway O stars) are expected to create cone-shaped bubbles, which were not seen by \citet{2006ApJ...649..759C}. Planetary nebulae can exhibit bipolar natures, but these are likely caused by close binary systems \citep{1987AJ.....94..671B,2006PASP..118..260S} or stellar magnetic fields \citep{2005A&A...432..273J}, but not by external magnetic fields \citep{2005AIPC..804...89S,2007MNRAS.376..378S}. A high-mass stellar bipolar outflow typically only exists inside the star's dense natal material \citep[but see][for parsec-long flows from lower-mass stars]{2004A&A...415..189M} and can be readily identified \citep[e.g.,][]{2009ApJS..181..360C}. Based on {\it Herschel Space Telescope} column density maps, \citet{2012A&A...542A..10A} found that Galactic bubble interiors have low densities. If stellar driven, they have evolved beyond the earliest stages of star formation in which bipolar outflows typically exist.

Magnetic fields may affect the evolution of Galactic bubbles via the fields providing anisotropic pressure, since it is easier for charged particles to move along field lines (i.e., lower pressure) than perpendicular to field lines (i.e., higher pressure). Only charged species sense magnetic fields, but they are collisionally coupled, to varying degrees, to the neutral species. Therefore, the presence of ordered magnetic fields (ordered on scales larger than the bubble diameter) can allow bulk flows along the field while inhibiting such flows along orthogonal directions. In the presence of an ordered, external magnetic field, uniform expansion of a Galactic bubble should be warped into an ellipsoid \citep{1992PASJ...44..177T}.

The outward flow of material making up the bubble must be driven. Possible mechanisms include combined stellar winds \citep{2006ApJ...649..759C}, supernovae \citep{2006AJ....131.2525H}, or expanding {\rm H\,{\scriptsize II}} regions \citep{2011ApJS..194...32A}. Since the evolution of bubbles likely depends on their energy sources, this work aims to avoid such dependence by a narrow focus on {\rm H\,{\scriptsize II}} region driven bubbles.

Starlight polarimetry, which traces magnetic fields via dust alignment, toward each Galactic bubble was drawn from early access to the Galactic Plane Infrared Polarization Survey \citep[GPIPS;][]{GPIPS_I}. Initial attempts to directly probe the magnetic fields of individual bubbles with GPIPS near-infrared (NIR) starlight polarimetry failed because the bubbles are predominantly beyond the distances readily probed by GPIPS. However, a statistical analysis of the relative alignment of a subsample of Galactic bubbles with the mean Galactic magnetic field orientation (which itself is predominantly parallel to the Galactic plane) does show evidence for correlation, leading to a possible {\rm H\,{\scriptsize II}} region evolutionary sequence. More turbulent (possibly younger) {\rm H\,{\scriptsize II}} region driven, elongated bubbles are better aligned with the Galactic plane than less turbulent (possibly older) {\rm H\,{\scriptsize II}} region bubbles. This observationally-driven result supports a scenario for the evolution of expanding {\rm H\,{\scriptsize II}} region driven bubbles into an external magnetic field.

In Section 2, the properties of {\rm H\,{\scriptsize II}} regions coincident with Galactic bubbles are summarized. The method used to measure Galactic bubble orientations is presented in Section 3 along with the orientations found. In Section 4, starlight polarimetry from GPIPS is used to estimate the mean magnetic field orientation of the Galaxy. Section 5 compares {\rm H\,{\scriptsize II}} region properties with the relative alignment of the elongated bubbles and external magnetic fields. A notional explanation for these observations is presented in Section 6 and conclusions are presented in Section 7.

\section{H {\scriptsize II} Regions Associated with Galactic Bubbles}
\label{4:datasets}

The Green Bank Telescope {\rm H\,{\scriptsize II}} Region Discovery Survey \citep[HRDS;][]{2010ApJ...718L.106B,2011ApJS..194...32A} is a targeted, 3 cm wavelength hydrogen radio recombination line (RRL) and continuum survey of the inner Galactic plane ($343\degr<\ell<67\degr$, $|b|<1\degr$) that overlaps with GPIPS and the Galactic Legacy Infrared Mid-Plane Survey Extraordinaire \citep[GLIMPSE;][]{2003PASP..115..953B}. HRDS targets were selected by searching for overlapping 24 $\mu$m emission from the {\it Spitzer Space Telescope} MIPS Galactic Plane Survey \citep[MIPSGAL; ][]{2009PASP..121...76C} and coincident 20 cm continuum emission from either the NRAO VLA Galactic Plane Survey \citep[VGPS; ][]{2006AJ....132.1158S} or NRAO VLA Sky Survey \citep[NVSS;][]{1998AJ....115.1693C}. These two target selection wavelengths (24 $\mu$m and 20 cm) were chosen because dust grains in {\rm H\,{\scriptsize II}} regions absorb stellar UV photons, which reemit at thermal infrared wavelengths, while the plasma surrounding the central ionizing star produces free-free thermal emission at centimeter wavelengths. HRDS simultaneously measured and averaged seven hydrogen recombination line profiles (H 86 $\alpha$ to H 93 $\alpha$, but excluding H 86 $\alpha$ because of confusion with higher order RRL transitions) towards each target \citep{2006AJ....132.2326B, 2010ApJ...718L.106B}. As of this writing, 603 discrete recomination line components from 448 lines of sight have been detected, and their radio properties cataloged \citep{2011ApJS..194...32A}.

\citet{2011ApJS..194...32A} compared the 603 HRDS RRL components with the 134 \citet{2006ApJ...649..759C} Galactic bubbles cataloged in the Northern Galactic plane. Thirty-three HRDS RRL components with coincident with twenty-seven Galactic bubbles (four bubbles have two coincident components and one bubble has three coincident components). Thus, the Galactic bubbles are assumed to trace the perimeters of {\rm H\,{\scriptsize II}} regions expanding into the interstellar medium. These twenty-seven Galactic bubbles constitute the subsample which will be studied in detail in later sections.

Selected data from HRDS for the thirty-three RRL components coincident with Galactic bubbles are listed in Table \ref{HII_properties}. The observed radial velocity, RRL linewidth, and peak RRL flux towards each of these Galactic bubbles are listed in columns 2, 3, and 4, respectively, in the Table. Several sightlines contain multiple RRL components, which are uniquely identified (e.g., N23a and N23b) in column 1 of the Table. As seen in the third column, several objects show large linewidths. An electron temperature of $10^4$ K results in a hydrogen thermal linewidth of 22 km~s$^{-1}$ \citep{1978ARA&A..16..445B}, hence large linewidths are suprathermal.

\begin{deluxetable*}{ccccccccc}
	\centering
	\tablewidth{0pt}
	\tablecaption{Galactic Bubble H {\scriptsize II} Region Properties\label{HII_properties}}
	\tablehead{   					 \colhead{Bubble} &
					 \colhead{$\rm RV_{\rm lsr}$\tablenotemark{1}} &
					 \colhead{Linewidth\tablenotemark{1}} &
					 \colhead{Peak H Line Flux\tablenotemark{1}} &
					 \colhead{Dist.} &
					 \colhead{Luminosity} &
					 \colhead{a} &
					 \colhead{b} &
					 \colhead{Spectral} \\
					 \colhead{Name} &
					 \colhead{(km~s$^{-1}$)} &
					 \colhead{(km~s$^{-1}$)} &
					 \colhead{(mJy)} &
					 \colhead{(kpc)} &
					 \colhead{(Jy kpc$^2$)} &
					 \colhead{(pc)} &
					 \colhead{(pc)} &
					 \colhead{Type} \\
					 \colhead{(1)} &
					 \colhead{(2)} &
					 \colhead{(3)} &
					 \colhead{(4)} &
					 \colhead{(5)} &
					 \colhead{(6)} &
					 \colhead{(7)} &
					 \colhead{(8)} &
					 \colhead{(9)}}
		\startdata
		\input{hii_table2.txt}
		\enddata
		\tablenotetext{1}{Data taken from \citet{2011ApJS..194...32A}.}
\end{deluxetable*}

Using the published HRDS line-of-sight velocity for each of the thirty-three recombination line components coincident with Galactic bubbles, kinematic distances were calculated using the \citet{1985ApJ...295..422C} rotation curve. All {\rm H\,{\scriptsize II}} regions were assumed to be located at the far kinematic distance or the tangent point, where appropriate, as reasoned below. An intrinsic velocity dispersion associated with random cloud motions in the Galaxy was also included. Though \citet{1985ApJ...295..422C} reports a cloud-cloud velocity dispersion of 3 km~s$^{-1}$, the more conservative value of 5 km~s$^{-1}$ \citep{1976ARA&A..14..275B} was adopted. The published uncertainty in the line radial velocity was added in quadrature to the random cloud dispersion to estimate upper and lower bounds for each kinematic distance. This propagated uncertainty was added and subtracted from each line velocity ($\rm RV_{\rm LSR}\pm\sigma_{\rm RV_{\rm LSR}}$ from column 2 in Table \ref{HII_properties}) and used to calculate upper and lower uncertainty bounds on each kinematic distance. The mean difference between these distance bounds and the central kinematic distance was adopted as the kinematic distance uncertainty (column 5 in Table \ref{HII_properties}).

The goal of HRDS was to improve the census of Galactic {\rm H\,{\scriptsize II}} regions, especially at large distances. Since the nearby population of {\rm H\,{\scriptsize II}} regions is well known, HRDS ignored them. This study was intentionally restricted to the HRDS data set because of the uniformity of their measurements of the hydrogen RRLs. So, not all {\rm H\,{\scriptsize II}} regions are at the far kinematic distance, but there is a very high probability that all the {\rm H\,{\scriptsize II}} regions taken from HRDS are.

For seventeen of these thirty-three HRDS RRL components, \citet{2012ApJ...754...62A} used {\rm H\,{\scriptsize I}} emission/absorption to resolve the near/far distance ambiguity. All seventeen of the RRL components were either at the far kinematic distance or at the tanget point. The {\rm H\,{\scriptsize I}} spectra towards the other HRDS RRL components contained only weak {\rm H\,{\scriptsize I}} features and no conclusions could be drawn. This finding lends support to the assumption above that all of the thirty-three {\rm H\,{\scriptsize II}} region driven bubbles are at the far kinematic distance.

With the HRDS RRLs and the kinematic distances to the thirty-three Galactic bubbles, additional properties can be derived for these {\rm H\,{\scriptsize II}} region driven bubbles, including the intrinsic luminosities and physical sizes of the major and minor axes of the bubbles (columns 6, 7, and 8 in Table \ref{HII_properties}).

The spectral type of the massive star creating each {\rm H\,{\scriptsize II}} region driven bubbles was estimated from the total RRL flux and the kinematic distance estimate for each RRL, using the Lyman continuum photon emission rates from \citet{2003ApJ...599.1333S}. \citet{2011ApJS..194...32A} predicted the expected HRDS RRL flux density from {\rm H\,{\scriptsize II}} regions hosting stars of these spectral types for a range of distances (0 - 23 kpc; shown as their Figure 12), assuming that a single star is responsible for all of the ionizing flux. From the previously determined peak line intensity (in mJy) and full-width at half-maximum (FWHM), each RRL component was integrated, assuming a Gaussian line profile. These spectral type estimates are listed in column 9 in Table \ref{HII_properties}.

\section{Galactic Bubble Orientations}
\label{4:bubble_orien}

\begin{figure*}[ht]
	\centering
		\includegraphics[angle=90,scale=0.8]{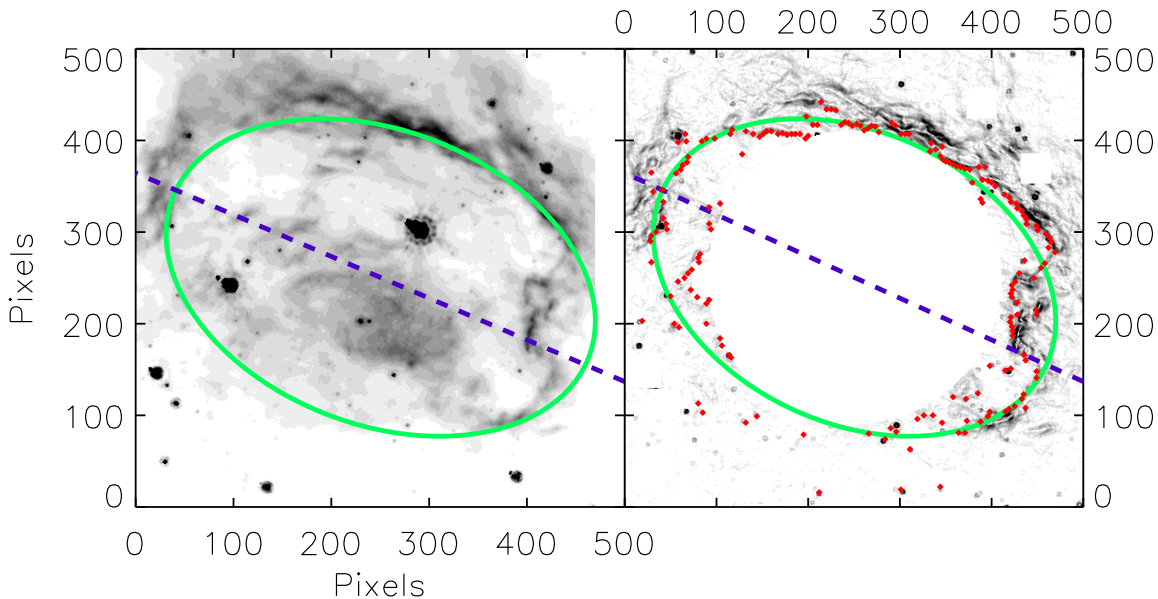}
	\caption{\label{ellipse_fitting}(Left panel) Greyscale GLIMPSE 8 $\mu$m image of bubble N21 with the final ellipse fit (ellipse in both panels) and the direction of the bubble position angle (diagonal dashed line in both panels). (Right panel) Greyscale GLIMPSE 8 $\mu$m image of N21 after application of a Sobel edge enhancement operator \citep{Sobel:1978:NCB}. Points indicate the first 2$\sigma$ datum along each of the 250 radial rays from the bubble's center that were used for fitting the ellipse.}
\end{figure*}

\citet{2006ApJ...649..759C} fit ellipses to their visually identified Galactic bubble sample, as seen in $8\,\mu$m GLIMPSE images. However, the position angles of the bubble major axes (`B-GPA,' in Galactic coordinates) were not retained. This key bubble parameter was needed for comparison of bubble orientations to the direction of the Galactic plane, and ultimately to the average projected Galactic magnetic field orientation.

Therefore, an algorithm was developed for fitting B-GPAs from GLIMPSE 8 $\mu$m images. Its action is illustrated for the bubble N21 in Figure \ref{ellipse_fitting}. In the left panel, the 8 $\mu$m emission from GLIMPSE (greyscale) is shown, in Galactic coordinates. The goal of the algorithm was to fit the inner ellipse parameters reported by \citet{2006ApJ...649..759C} to the 8 $\mu$m emission on the inner boundaries of each bubble and so recover the B-GPAs.

First, bright point sources in each image were identified with the DAOPHOT FIND routine \citep{1987PASP...99..191S} and masked, using 18$\times$18 arcsec squares. A \citet{Sobel:1978:NCB} edge enhancement operator was applied to the images. This operator calculates the horizontal ($\textbf{G}_x$) and vertical ($\textbf{G}_y$) intensity gradients across each pixel and returns an approximation of the amplitude of the total intensity gradient (\textbf{G}). In practice, these gradient images are generated by convolving ($\otimes$) the image (\textbf{A}) with two different kernels:

\begin{equation}
	\textbf{G}_x = \left[ 
		\begin{array}{ccc}
		-1 & 0 & +1 \\
		-2 & 0 & +2 \\
		-1 & 0 & +1 
		\end{array} \right] \otimes \textbf{A}\,;\,\,
\textbf{G}_y = \left[ \begin{array}{ccc}
	-1 & -2 & -1 \\
	 0 & 0  &  0 \\
	+1 & +2 & +1 \end{array} \right] \otimes \textbf{A}.
\end{equation}

These two resulting images are combined to create the full gradient image $\textbf{G} = \sqrt{\textbf{G}_x^2 + \textbf{G}_y^2}$. Original image ($\textbf{A}$) regions of constant flux return zero gradients.

Circular regions centered at the reported bubble centers, with radii equal to the reported inner minor axes, were masked in each resulting Sobel image (${\textbf G}$). The resulting Sobel image for bubble N21 is shown in the right panel of Figure \ref{ellipse_fitting}.

Vectors of pixel values were extracted from the masked Sobel image along 250 equally spaced (in azimuthal angle) radial rays from the reported bubble center. The means and standard deviations of values contained in each data vector were found. Starting from the center of the image and moving radially outward, the location of the first datum exhibiting more than a two sigma deviation above the average value for that vector was identified. These 250 locations are shown by the diamond symbols in the right panel of Figure \ref{ellipse_fitting}. These ($\ell$, $b$) locations were the basis for new ellipse fitting.

The routine MPFITELLIPSE from the MPFIT package \citep{2009ASPC..411..251M} was used to fit one ellipse to the locations of the 250 points described above for each bubble. The center coordinates, major axis, and minor axis were all fixed and taken from \citet{2006ApJ...649..759C}, leaving B-GPA as the only fit parameter. The result of the fit for bubble N21 is shown as the ellipses in both panels of Figure \ref{ellipse_fitting}. The B-GPA is shown as dashed lines in the Figure.

\begin{figure}
	\centering
		\plotone{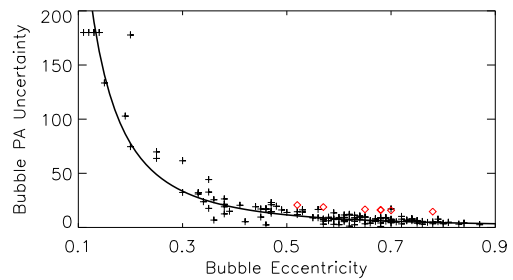}
	\caption{\label{bubble_pa_err}The dependence of fitted Galactic bubble position angle uncertainty on bubble eccentricity. The solid line is a power law fit to the data. Seven bubbles were fit by hand and their estimated uncertainties are shown by the diamonds. Two of the seven bubbles had very similar eccentricities and their symbols overlap.}
\end{figure}

This procedure was applied to all 134 bubbles in the Northern Galactic plane reported by \citet{2006ApJ...649..759C}. The resulting B-GPAs are listed in Table \ref{bubble_parms}, along with their uncertainties. The uncertainties in B-GPAs were found to depend on bubble eccentricity, as revealed in Figure \ref{bubble_pa_err}. For small eccentricities, bubbles shapes and position angles are more difficult to ascertain. For bubbles with very small eccentricities, the fitting returned B-GPA uncertainities of $180\degr$. The behavior was well-fit by a power law ($\sigma_{PA}=2.76\, {\rm ecc}^{-2.07}$), based on an F-test, and the resulting fit is shown by the solid line in Figure \ref{bubble_pa_err}. This same behavior was recovered when only the subsample of thirty-three {\rm H\,{\scriptsize II}} region driven bubbles was considered.

The presence of very bright stars in the GLIMPSE field-of-view as well as foreground infrared dark clouds caused the fitting procedure to fail for seven bubbles, which are identified in Table \ref{bubble_parms}. For these bubbles, B-GPAs were fit by hand and the uncertainties (diamonds) were conservatively estimated using the published bubble eccentricities and the power law fit (solid line), with an additional $10\degr$ of uncertainty.

The bubble identifiers, eccentricities, and the major axes of the inner bubble wall \citep[all from][]{2006ApJ...649..759C} for all 134 northern Galactic bubbles are listed in columns 1, 2, and 3 of Table \ref{bubble_parms}. The B-GPAs and their uncertainties are listed in column 4 of that Table.

\section{Starlight Polarimetry Towards Galactic Bubbles}

All available early access near-infrared starlight polarimetry towards towards Galactic bubbles was obtained from GPIPS. GPIPS is an H-band (1.6 $\mu$m) linear imaging polarimetry survey of the northern Galactic plane ($18\degr<\ell<56\degr$, $|b|<1\degr$) using the Mimir instrument \citep{2007PASP..119.1385C} on the 1.8 m Perkins Telescope outside of Flagstaff, AZ. GPIPS observations and data reduction pipelines are discussed in detail in \citet{GPIPS_II,GPIPS_I}. The data used here were observed between 2007 May and 2011 June. While most Galactic bubbles were observed, significant gaps in coverage surrounding many of them remained at the time of this analysis.

Originally, a goal was to measure stars through the interiors of bubbles, through regions exterior to bubbles, and (if possible) through the thick edges of bubbles. With this goal in mind, only bubbles with inner major axis extents greater than one arcminute were initially considered, because of the expected GPIPS polarimetric sky sampling rate of about one star per square arcminute. However, this angular sampling rate would not provide statistically significant samples of stars for most of the \citet{2006ApJ...649..759C} Galactic bubbles. Furthermore, the twenty-seven Galactic bubbles with distance estimates are typically beyond the GPIPS distance horizon of approximately 7 kpc \citep{GPIPS_I}, limiting the utility of starlight polarimetry to probe the magnetic fields around those objects.

Instead, to test the role of magnetic fields in bubble asymmetry, bubble orientations were compared with average magnetic field directions in the Galactic disk. This change in scope allowed study of all 134 \citet{2006ApJ...649..759C} bubbles in the Northern Galactic plane instead of only those bubbles with the largest angular sizes.

Lacking direct measures of the magnetic field orientations surrounding each bubble, the bubbles were instead assumed to be embedded in a Galactic-scale, ordered magnetic field. The disk of the Milky Way has a large-scale, toroidally-dominated magnetic field \citep[e.g.,][]{2000AJ....119..923H, 2011ApJ...728...97V, 2012ApJ...749...71P}. All of the Galactic bubbles in the \citet{2006ApJ...649..759C} sample are within one degree of the Galactic midplane, placing them well within the toroidally dominated zone. For any disk-symmetric Galactic magnetic field, the projection of the toroidal component will exhibit NIR background starlight polarization position angles (`P-GPA'), in Galactic coordinates, of about $90\degr$, for nearly all Galactic longitudes and distances \citep{2011ApJ...740...21P}.

This approximation breaks down for lines-of-sight along the local magnetic field direction. In the Solar neighborhood, this occurs at Galactic longitudes $\ell=90\degr+p$ and $\ell=270\degr+p$, where p is the magnetic pitch angle. \citet{2012ApJ...749...71P} found $p=-6\pm2\degr$, similar to values found by other studies using a variety of techniques \citep{1988AJ.....95..750V, 1994A&A...288..759H, 1996ApJ...462..316H, 2007EAS....23...19B, 2011ApJ...738..192P}. For this magnetic pitch angle, the longitudes of vanishing polarization become $\ell=84\degr$ and $264\degr$, well outside the longitude range of GPIPS and where these Galactic bubbles are located.

Weaker vertical and radial magnetic field components \citep[e.g.,][]{2000AA...358..125F} will perturb the dominant toroidal magnetic field. These perturbations will cause P-GPA to differ from perfect alignment with the Galactic plane \citep[see example deviation in][]{GPIPS_III}. Therefore, GPIPS starlight polarimetry was used to characterize the mean P-GPA for the Galactic magnetic field within several kpc of the Sun, and this characterization was assumed to apply to all locations within the Galaxy.

Towards each Galactic bubble, all available GPIPS starlight polarimetry data within three times the outer major axis radius ($R_{\rm out}$) of each bubble center, as reported by \citet{2006ApJ...649..759C}, were collected. These starlight polarization position angles were combined into one weighted mean P-GPA around each bubble, and this value is listed in column 8 of Table \ref{bubble_parms}. To keep a few very low uncertainty measurements from dominating the weighted mean, all starlight polarization position angle uncertainties were floored to $5\degr$ or greater.

\begin{figure}
	\centering
		\plotone{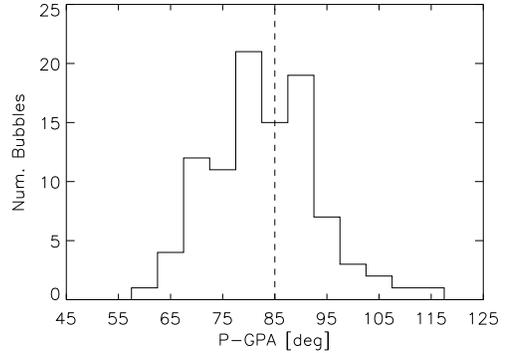}
	\caption{\label{gpa_los}Histogram of the weighted average GPIPS polarization position angles (P-GPA) towards each of the 97 bubbles with GPIPS polarimetry. The vertical dashed line at $85.3\degr$ shows the unweighted average of these measurements. The sharpness of the distribution suggests that P-GPA$=85\degr$ is representative of average orientation of the projected Galactic magnetic field in the disk.}
\end{figure}

The unweighted average starlight polarization $<$P-GPA$>$ towards all of the bubbles, listed in column 8 in Table \ref{bubble_parms}, is $85.3\degr$ with a dispersion of $10\degr$. An F-test showed that assuming a constant value for the magnetic field orientation across this longitude range was appropriate and no higher-order longitude-dependent terms were needed. Thirty-six Galactic bubbles, N1-N19 and N123-N134, are outside of the GPIPS region. Therefore, only 97 bubbles had GPIPS measurements, and the histogram of these $<$P-GPA$>$ is shown in Fig. \ref{gpa_los}.

This mean value agrees with thoretical predictions \citep[e.g., ][]{1993AARv...4..449W, 2000AA...358..125F, 2010AA...512A..61M} and previous observations \citep[e.g., ][]{1970MmRAS..74..139M, 2000AJ....119..923H, 2008AA...486..819M}. Analysis of NIR starlight polarimetry in the outer \citep{2012ApJ...749...71P} and inner \citep{GPIPS_III} Galaxy, also supports the conclusion that a single mean orientation is a reasonable characterization of the Galactic magnetic field in the disk. 

\section{Analysis}
\label{4:discussion}

Using the measured bubble properties from Section \ref{4:bubble_orien} and the assumption of a constant Galactic magnetic field orientation, the absolute difference between the bubble orientation and the average projected Galactic magnetic field orientation ($\Delta {\rm GPA} = |$B-GPA$-$P-GPA$|$) can be used to test whether magnetic fields can explain the observed bubble eccentricities. Two limiting cases are: (1) magnetic fields completely determine the bubble orientation, or, (2) magnetic fields have no effect on the bubble orientation. If the former, then the $\Delta {\rm GPA}$ histogram should be sharply peaked around $\Delta {\rm GPA}=0\degr$. If magnetic fields have no effect on bubble orientation, the histogram of $\Delta {\rm GPA}$ should be flat, consistent with random bubble orientations.

\begin{figure}
	\centering
		\plotone{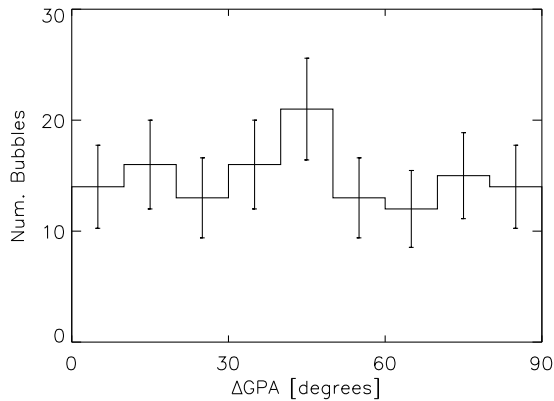}
	\caption{\label{delta_gpa_bg}Histogram of the absolute difference between the bubble major axis orientation (B-GPA) and the average projected Galactic magnetic field orientation (P-GPA=$85\degr$).}
\end{figure}

Figure \ref{delta_gpa_bg} shows the histogram of $\Delta {\rm GPA}$ for the 134 Northern Galactic bubbles: the distribution appears to be flat. The cumulative distribution function (CDF) of the $\Delta {\rm GPA}$ data was compared to the CDF of a flat distribution, using a KS test, to find a 91\% chance that the $\Delta {\rm GPA}$ data were drawn from a flat distribution. This fails to reject the null hypothesis of no preferential alignment between the average Galactic magnetic field orientations and bubble elongations. When the bubbles are considered as a single populations, bubble major axes show no preferential alignment with the average Galactic magnetic field orientation, and bubble eccentricity is not caused by external magnetic fields. $\Delta {\rm GPA}$ was also examined against estimated spectral type of the star powering each {\rm H\,{\scriptsize II}} region, and no correlation ($R^2=0.0015$) was found. Yet when subsamples of bubbles were created, correlations emerged, as next described.

\subsection{An H {\scriptsize II} Region Magnetic Evolutionary Sequence}
\label{HII_evol}

Several of the hydrogen RRL components listed in Table \ref{HII_properties} are suprathermal suggesting that additional line broadening mechansims (e.g., pressure broadening, large scale systematic motions, and turbulence) are important. At the frequencies of the RRL observations (9 GHz), and assuming an electron density of $10^3$ cm$^{-3}$ at a temperature of $10^4$ K in the line emitting regions, pressure broadening should be 1.3\% of the thermal broadening \citep[][and references therein]{2008ApJ...672..423K}, therefore pressure broadening can be ignored. The effects of large scale systematic motions and turbulent motions in the gas cannot be disentangled with these observations, so their combined effects on the RRL linewidths will be considered in the following analysis and collectively refered to as turbulence. Assuming a temperature of $10^4$ K in the line emitting regions, the \citet{2011ApJS..194...32A} RRL linewidths can be decomposed into thermal and turbulent components.

In Kolmogorov-like turbulence \citep{1941DoSSR..30..301K}, turbulent energy is injected into a system at some characteristic length scale. The energy is redistributed to smaller size scales, reaching a ``Kolmogorov'' energy spectrum of $E(k) \propto k^{-5/3}$, though no assumption is made about the actual spectral indicies in the {\rm H\,{\scriptsize II}} regions probed here. In the system, energy is dissipated at some small, resistive, length scale. There is a monotonic flow of energy from larger to smaller length scales while maintaining the turbulent energy spectrum. \citet{2012ApJ...750L..31R} considered the effects of an expanding or contracting volume containing a turbulent gas. Their simulations showed that an expanding turbulent gas will experience `adiabatic cooling' and that the amplitude of the turbulent energy spectrum will decrease as the volume increases. Under these conditions, the turbulent velocity components should decrease with time, while the $10^4$ K thermal linewidth component is unchanged. Perhaps suprathermal linewidths are associated with younger, turbulent {\rm H\,{\scriptsize II}} regions and smaller linewidths are associated with older, less turbulent {\rm H\,{\scriptsize II}} regions.

\begin{figure}
	\centering
		\plotone{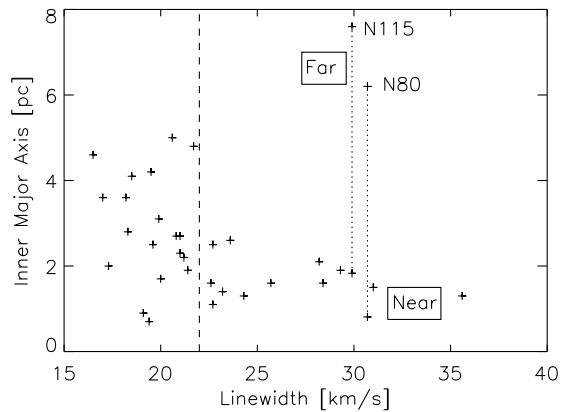}
	\caption{\label{size_linewidth}Measured inner major axis for bubbles along 33 lines-of-sight with HRDS recombination line observations as a function of hydrogen radio recombination linewidth. The vertical dashed line at 22 km~s$^{-1}$ represents the thermal linewidth expected for a $10^4$ K gas. The major axes of bubbles N115 and N80, assumed to be at their far kinematic distances, were also calculated for their near kinematic distances, the two estimates are connected by the dotted lines.}
\end{figure}

Figure \ref{size_linewidth} plots the physical size of the thirty-three {\rm H\,{\scriptsize II}} region driven bubbles (derived from the kinematic distance estimates) against the HRDS linewidths of those {\rm H\,{\scriptsize II}} regions. A few interesting features are seen in Figure \ref{size_linewidth}. The data seem to break into two groups at a linewidth of approximately 22 km~s$^{-1}$ (shown by the vertical dashed line). At smaller linewidths, a wider range of physical {\rm H\,{\scriptsize II}} region major axes is seen; at larger linewidths, the distribution of major axes is generally narrower and at smaller physical sizes. Figure \ref{size_linewidth} shows that radius is not an indicator of linewidth. Only physically smaller {\rm H\,{\scriptsize II}} regions (except where there is a distance ambiguity; e.g., N80 and N115) exhibit suprathermal linewidths. The full range of physical sizes is represented by the {\rm H\,{\scriptsize II}} regions showing subthermal linewidths. 

Physical size, by itself, should not be a good {\rm H\,{\scriptsize II}} region age indicator because the growth rate of an {\rm H\,{\scriptsize II}} region will be strongly influenced by the energy output (radiation and stellar wind) of the central star. Model stellar atmospheres by \citet{2003ApJ...599.1333S} show that the ionizing photon rate can vary by two orders of magnitude in going from O-type to early B-type stars. This causes any radius-age relation to become degenerate with the spectral type of the central star, in the sense that young {\rm H\,{\scriptsize II}} regions with massive stars will have the same radii as older {\rm H\,{\scriptsize II}} regions driven by less massive stars. While an estimate of the spectral type of the star in each of {\rm H\,{\scriptsize II}} region was made, this estimate assumed a single ionizing star was driving the {\rm H\,{\scriptsize II}} region. For the actual {\rm H\,{\scriptsize II}} regions, multiple massive stars could contribute to the ionizing fluxes and so would be expected to further confuse or mask any age-radius relation. 

To quantify any effect of spectral type on the other observed {\rm H\,{\scriptsize II}} region properties, a number was assigned to each spectral type (3 for O3, 4 for O4,...,10 for B0). The observed hydrogen linewidths and spectral types showed a weak linear correlation ($R^2=0.15$), with earlier-type stars showing slightly larger linewidths. The correlation is too weak for spectral type to account for the break seen at a linewidth of 22 km s$^{-1}$.

In Figure \ref{size_linewidth}, two outliers (bubbles N80 and N115) are seen in the high linewidth region, exhibiting inner major axes greater than 6 pc. Both of these bubbles were assumed to be at their far kinematic distances. \citet{2012ApJ...754...62A} used {\rm H\,{\scriptsize I}} spectra to confirm that N80 is at the far kinematic distance, but the {\rm H\,{\scriptsize I}} spectral quality was too low to break the distance ambiguity for N115. Their physical sizes were recalculated assuming they were instead at the near kinematic distance (1.47 kpc for N80 and 1.96 kpc for N115). If these bubbles are at their near kinematic distances, then they agree with the overall trend of two groups separated by a characteristic linewidth. Nevertheless, the far kinematic distances were adopted for the following analysis.

Since the temperature of an {\rm H\,{\scriptsize II}} region is expected to be close to $10^4$ K over most of its lifetime \citep{1978ARA&A..16..445B}, the larger linewidths are likely caused by turbulent motions (large scale, systematic motions and small scale turbulence) in the line emitting regions of each bubble. Also, the physical size of an {\rm H\,{\scriptsize II}} regions is a monotonically increasing function with time, which may reach an equilibrium size but should never decrease. Together, the small physical sizes and suprathermal linewidths suggest that the {\rm H\,{\scriptsize II}} regions with larger linewidths may be younger.

\begin{figure}
	\centering
		\plotone{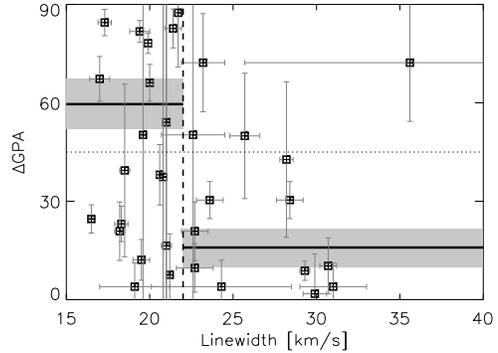}
	\caption{\label{deltapa_linewidth}Dependence of $\Delta$GPA on hydrogren radio recombination spectral profile linewidth. The vertical dashed line at 22 km~s$^{-1}$ represents the thermal linewidth expected for a $10^4$ K gas. The horizontal dotted line shows the expected $\Delta$GPA for random orientations. The thick horizontal lines show the weighted mean $\Delta$GPA in the subthermal and suprathermal zones. The grey regions represent the $1\sigma$ uncertainties on the means in each zone. As linewidths increase beyond 22 km~s$^{-1}$, there is generally better alignment between the bubble and magnetic position angles.}
\end{figure}

Figure \ref{deltapa_linewidth} shows $\Delta {\rm GPA}$ as a function of linewidth. The vertical dotted line drawn at 22 km~s$^{-1}$ shows the linewidth beyond which turbulence must be significant. Beyond 22 km~s$^{-1}$, there appears to be better alignment (low $\Delta {\rm GPA}$) between each bubble and the mean Galactic magnetic field.

Two subsamples were created and analyzed for relative alignment of the bubbles and mean magnetic field orientation; bubbles with subthermal linewidths ($\Delta {\rm v}\leq$ 22 km~s$^{-1}$) and suprathermal linewidths ($\Delta {\rm v}>$ 22 km~s$^{-1}$). For a uniform distribution of relative alignments ($\Delta {\rm GPA}$ independent of linewidth), the average value of $\Delta {\rm GPA}$ should be $45\degr$. The mean weighted $\Delta {\rm GPA}$ values for the subthermal and suprathermal subsamples linewidths are $60\degr\pm 8\degr$ and $16\degr\pm 6\degr$, respectively. The uncertainty on each average is the weighted dispersion in each region. Furthermore, a KS test shows only a 30\% chance that the $\Delta {\rm GPA}$ of the subthermal subsample and the suprathermal subsample are drawn from the same parent population. Therefore, Galactic bubbles with recombination linewidths greater than 22 km~s$^{-1}$ are better aligned with the average Galactic magnetic field orientation than Galactic bubbles with smaller linewidths.

While the suprathermal sample shows a significant ($\sim 5\sigma$) deviation from the expected mean $\Delta {\rm GPA}$ for a random distribution, the subthermal sample also shows a marginal deviation ($\sim 2\sigma$) from that same mean.

\section{Discussion}

The observational evidence presented above shows that external magnetic fields are important during the earliest phases of the evolution of {\rm H\,{\scriptsize II}} regions. Having shown this, a key following question is \textit{what external magnetic field strengths are required to generate such effects on the morphologies of the bubbles?}

\subsection{Magnetic-to-Ram Pressure Ratio}

A first estimate of the requsite magnetic field strengths can be found by equating the magnetic and ram pressures acting on an expanding ionization front:
\begin{equation}
	\eta\frac{1}{2}\rho_{ \rm shell}u_{\rm shell}^2 = \frac{B^2}{8\pi},
\end{equation}
where $\eta$ is an efficiency factor, $\rho_{\rm shell}$ is the density of the shell, $u_{\rm shell}$ is the expansion speed of the shell, and $B$ is the magnetic field strength at the outer boundary of the shell. An efficiency factor ($\eta$) is included because the ram and magnetic energy densities need not be equal for the magnetic field to perturb the evolution of the expanding front.

By Bernoulli's principle, a pressure differential in a fluid element will give rise to a velocity, $u \propto \sqrt{dp}$, so that magnetic pressure acts to a decrease the bubble's expansion velocity. In the limit of equal ram and magnetic pressures, expansion halts.

\begin{figure}[h]
	\plotone{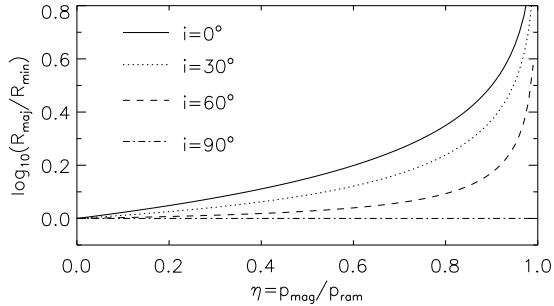}
	\centering
	\caption{\label{axis_predic}The predicted major-to-minor axis ratio for bubbles experiencing different magnetic-to-ram pressure ratios at different viewing angles. For $i=90\degr$, the effect is unobservable since the anisotropic force acts along the line-of-sight.}
\end{figure}

This near equivalence of differential pressure and velocity allows the relative pressures of the \citet{2006ApJ...649..759C} bubbles to be estimated. The ratio of the major and minor axes of each bubble will be equal to the ratio of its expansion velocities,
\begin{equation}
	\frac{R_{\rm maj}}{R_{\rm min}} = \frac{v_{\rm maj}}{v_{\rm min}}.
\end{equation}
Applying Bernoulli's principle and including projection effects, the observed bubble axis ratio becomes:
\begin{equation}
	\frac{R_{\rm maj}}{R_{\rm min}} = \sqrt{\frac{p_{\rm ram} - sin(i) \, p_{\rm mag}}{p_{\rm ram} - p_{\rm mag}}},
\end{equation}
where $i$ is the inclination angle of the major axis to the plane of the sky (e.g., $i=90\degr$ would be a line-of-sight along the poles of the bubble, which would appear as a circle in projection), $p_{\rm ram}$ is the expansion ram pressure, and $p_{\rm mag}$ is the magnetic pressure affecting expansion along the minor axes. The predicted effects of the magnetic-to-ram pressure ratio and inclination angle on the observed bubble axis ratio are shown in Figure \ref{axis_predic}. As the magnetic-to-ram pressure ratio increases, the projected bubbles are distorted from circles ($R_{\rm maj}/R_{\rm min}=1$) into ellipses. The inclination angle also affects the observed axis ratio. In the limit of $i=90\degr$, all bubbles appear as circles.

Since inclination will \textit{always} cause the observed bubble axis ratio to be smaller than the true axis ratio, bubbles with the largest observed axis ratios (least likely to suffer from strong projection effects) can be used to estimate the typical magnetic-to-ram pressure ratio. Table \ref{magnetic_ram} lists the major-to-minor axis ratios and the corresponding lower limits on the magnetic-to-ram pressure for the twelve bubbles with hydrogen recombination linewidths greater than 22 km~s$^{-1}$ (those showing the elongation-magnetic field correlation). As an example, the largest Galactic bubble axis ratio in this sample is 1.38 (bubble N80). Assuming that this bubble has an inclination angle of $0\degr$, then $p_{\rm mag} = 0.47 p_{\rm ram}$ (and the coefficient is equal to the $\eta$ efficiency term). If the inclination is non-zero, the magnetic-to-ram pressure ratio is even larger. Using the observed bubble axis ratios listed in Table \ref{magnetic_ram}, a lower limit on $\eta$ can be estimated as the unweighted mean of these estimates, $<\eta>=0.29$. This remains a lower limit, because projection effects will cause the observed axis ratios (and therefore $\eta$ values) to be biased towards unity, as shown in Figure \ref{axis_predic}, and so underestimated $\eta_{\rm true}$. Uncertainties on bubble major and minor axis radii were not reported by \citet{2006ApJ...649..759C}, so uncertainties in the axis and pressure ratios were not propagated.

\begin{deluxetable}{ccc}
	\tabletypesize{\footnotesize}
	\tablecaption{\label{magnetic_ram}Magnetic-to-ram Pressure Estimates}
	\tablewidth{0pt}
	\tablecolumns{3}
	\centering
	\tablehead{\colhead{Bubble} &
						 \colhead{Axis} &
						 \colhead{Pressure} \\
						 \colhead{Name} &
						 \colhead{Ratio} &
						 \colhead{Ratio $\eta$}\\
						 \colhead{(1)} &
						 \colhead{(2)} &
						 \colhead{(3)} \\
						 }
	\startdata
		N23 & 1.08 & 0.14 \\
		N27 & 1.20 & 0.31 \\
		N31 & 1.30 & 0.41 \\
		N50 & 1.13 & 0.21 \\
		N53 & 1.01 & 0.02 \\
		N31 & 1.22 & 0.33 \\
		N66 & 1.28 & 0.39 \\
		N73 & 1.06 & 0.12 \\
		N80 & 1.38 & 0.47 \\
		N96 & 1.23 & 0.34 \\
		N110 & 1.09 & 0.15 \\
		N115 & 1.27 & 0.38 \\
	\enddata
\end{deluxetable}

\subsection{Magnetic Field Strength Estimates}

A lower limit to the mass of an expanding shell is the equivalent mass of the ambient interstellar medium swept up by the shell. The shell's density can be estimated as
\begin{equation}
	\label{rho}
	\frac{\rho_{\rm shell}}{\rho_{\rm ISM}} = \frac{1}{{1 - [(1-<T>)]^3}}
\end{equation}
where $\rho_{\rm ISM}$ is the average mass density of the interstellar medium, and $<T>$ is the average fractional thickness of the shell, calculated by \citet{2006ApJ...649..759C}. For the twelve bubbles with large hydrogen recombination linewidths ($\Delta {\rm v} >22$ km~s$^{-1}$), the average shell thicknesses is $<T>=0.24$, and $\rho_{\rm shell}=1.78\rho_{\rm ISM}$.

Assuming the shell is expanding at $\sim 10$ km~s$^{-1}$, the ion sound speed of a 10$^4$ K gas, into a medium with a density equal to that of the diffuse interstellar medium \citep[1.4 amu cm$^{-3}$;][]{2009ApJ...703.1352K}, the surrounding magnetic field must be 72 $\mu$G to exert a magnetic pressure equal to the shell's ram pressure. Using the pressure efficiency factor found earlier ($\eta=0.29$), this decreases to 21 $\mu$G.

This neglects the mass of any denser natal cloud that might also be swept up into the expanding shell, which would increase the shell mass density. Assume an O4 main sequence star, with mass of $\sim60$ $M_\sun$, was at the center of a 1 pc radius expanding {\rm H\,{\scriptsize II}} region. If the star formation efficiency is 30\% \citep{2011ApJ...729..133M}, there would be an additional 200 $M_\sun$ of natal material in the shell. In this case, the magnetic field must be amplified to 2.8 mG to equal the ram pressure (810 $\mu$G including $\eta$). For a 25 $M_\sun$ O8 main sequence star, this becomes 1.8 mG (520 $\mu$G).

This notion may be supported by OH maser observations towards {\rm H\,{\scriptsize II}} regions. OH masers often arise in the shocked neutral gas surrounding ultracompact {\rm H\,{\scriptsize II}} regions and the magnetic field strength can be measured via Zeeman splitting \citep{2006ApJS..164...99F}. \citet{1998A&A...340..521D} observed magnetic field strengths of 4 to 6 mG near the ultra-compact {\rm H\,{\scriptsize II}} regions ON1 and W51. \citet{2005ApJS..160..220F} surveyed 18 Galactic massive star-forming regions at 1665 MHz with the Very Large Baseline Array and observed mG strength fields in 184 unique OH masers. Methanol masers have also been used to probed magnetic field strengths in high-mass star-forming regions. \citet{2008A&A...484..773V} find an average magnetic field strength of 23 mG from 24 bright methanol masers. More recently, \citet{2012arXiv1207.3550G} has found mG magnetic fields towards OH masers in high-mass star forming regions towards the Carina-Sagittarius spiral arm tangent.

These maser measurements are toward very young {\rm H\,{\scriptsize II}} regions, much smaller and younger than the {\rm H\,{\scriptsize II}} regions interior to Galactic bubbles. However, future high resolution studies of their morphologies in comparison to the local magnetic field orientation (perhaps through high resolution polarized dust emission at submillimeter wavelengths) would provide additional insight into the magnetic field properties of {\rm H\,{\scriptsize II}} region driven Galactic bubbles.

Detailed calculation of the ram pressure for each bubble is not possible here. For isolated high mass stars, that would require either spectroscopic classification of the central ionizing star or direct measurement of the mass (i.e., gas column for atomic or molecular species at radio wavelengths) and velocity of the expanding shell. Spectral classifications are not available for these stars, and the assumption of a single, ionizing star may not apply if several early-type stars contribute ionizing radiation. High-resolution spectroscopic observations with airborne instruments, such as GREAT \citep{2012A&A...542L...1H} on SOFIA, may allow more reliable measurements of, for example, the OI (63 $\mu$m) traced mass and velocity structure around {\rm H\,{\scriptsize II}} regions.

The average magnetic field strength in the solar neighborhood is around 6 $\mu$G \citep{2009ASTRA...5...43B}. In the limit of flux-freezing, and assuming that a magnetic field of this strength threads all of the material swept up by a bubble shell, the O4 case would need to sweep up a volume having a cross-sectional area equal to $2800/6\approx470$ times its current area for the magnetic pressure to equal the ram pressure. For a shell of cross-sectional area $A=2 \pi r dr$, and assuming that $dr=0.2r$, this is an increase in the radius of the shell by a factor of 21. In expanding from a radius of 0.05 pc to 1 pc, the shell will accumulate enough magnetic flux for the magnetic pressure to equal the ram pressure. As discussed earlier, the magnetic pressure is typically only a fraction of the ram pressure, decreasing the expansion factor necessary to affect the bubble's evolution. However, relaxing the flux-freezing requirement would allow magnetic flux to diffuse into the bubble and (if expanding at sub-alfv\'{e}nic speeds) possibly away from the expanding shell.

As shown in Figure \ref{deltapa_linewidth} and discussed in Sec. \ref{HII_evol}, this magnetic alignment mechanism must end by the time the recombination linewidth falls to 22 km~s$^{-1}$. This roughly corresponds to when the inner major axis has expanded to approximately 2 pc (see Figure \ref{size_linewidth}). Magnetic warping of bubble shells could also be less important for high mass star formation in a dense, extended medium where mass loading of the expanding shell causes it to slow before appreciable magnetic amplification can occur.

\section{Conclusions}

The interaction of {\rm H\,{\scriptsize II}} region driven Galactic bubbles with the Galactic magnetic field has been examined. Starlight polarimetry from GPIPS was used to probe the large-scale properties of the Galactic disk magnetic field and revealed that the projected orientation of the Galactic magnetic field is approximately constant ($\langle$P-GPA$\rangle =85\degr\pm10\degr$) over the Galactic longitude range $18\degr<\ell<56\degr$, and consistent with previous work showing the toroidally-dominated nature of the Galactic magnetic field in the disk. Existing hydrogen radio recombination line data from HRDS were used to measure physical properties of {\rm H\,{\scriptsize II}} region driven bubbles and to estimate their kinematic distances.

{\rm H\,{\scriptsize II}} region driven bubbles with suprathermal ($>$ 22 km~s$^{-1}$) hydrogen recombination linewidths are preferentially aligned with the average orientation of the Galactic magnetic field in the disk. {\rm H\,{\scriptsize II}} region driven bubbles with subthermal ($<$ 22 km~s$^{-1}$) linewidths are consistent with random alignments. Recombination linewidths were also shown to be anticorrelated with the physical sizes of bubbles. Since {\rm H\,{\scriptsize II}} region turbulence should decrease with time, recombination linewidths may be used as crude age indicators for {\rm H\,{\scriptsize II}} regions. The spectral types of the stars at the centers of the {\rm H\,{\scriptsize II}} regions were estimated from the HRDS kinematic distances and RRL flux densities. Spectral type has a weak effect on the observed linewidth (but is unable to account for the break at 22 km~s$^{-1}$), and showed no correlation with the relative alignment between bubble long axes and the mean Galactic magnetic field direction ($\Delta {\rm GPA}$).

The major-to-minor axis ratios of the Galactic bubbles from \citet{2006ApJ...649..759C} were used to estimate the magnetic-to-ram pressure ratio for each bubble. A lower limit to this ratio was shown to be 0.29 for the sample of magnetically-aligned bubbles (linewidths $>$ 22 km~s$^{-1}$).

These findings have led to development of a scenario for the evolution of the relative alignment between {\rm H\,{\scriptsize II}} region driven bubbles and the Galactic magnetic field. New {\rm H\,{\scriptsize II}} regions are small and characterized by large turbulent energy densities, showing suprathermal linewidths. As {\rm H\,{\scriptsize II}} regions grow and age, their ionization fronts expand into magnetized interstellar media, where the shells interact with the large-scale Galactic magnetic field. The magnetic field is entrained in the shells and becomes amplified. The increasing magnetic field strength inhibits expansion perpendicular to the large-scale field orientation, causing the bubble to expand preferentially along the external magnetic field direction. This creates elliptical bubbles preferentially aligned with the magnetic field direction.

As the {\rm H\,{\scriptsize II}} regions further age, the shells slow because of mass loading and the magnetic field weakens via magnetic diffusion. Around this time, turbulent energy in the hydrogen recombination line emitting region dissipates, causing the linewidths to decrease to thermal values. Magnetic fields no longer impress a preferential orientation on the {\rm H\,{\scriptsize II}} region's continued expansion and other forces (e.g., local gas density variations) come to dominate the morphological evolution of the {\rm H\,{\scriptsize II}} region.

\acknowledgements

The authors would like to thank L. Anderson and T. Bania for useful conversations about HRDS. This research used data from the Boston University (BU) Galactic Plane Infrared Polarization Survey (GPIPS), funded in part by NSF grants AST 06-07500 and 09-07790. GPIPS used the Mimir instrument, jointly developed at BU and Lowell Observatory and supported by NASA, NSF, and the W.M. Keck Foundation.

\bibliography{PavelBib_bubbles.bib}

\begin{longtable}{ccccc}
	\tablecaption{Bubble Parameters\label{bubble_parms}}
	\centering
	\tablewidth{0pt}
	\tablecolumns{5}
	\tablehead{
   \colhead{Name\tablenotemark{1}} &
   \colhead{Ecc.\tablenotemark{1}} &
   \colhead{$R_{\rm maj}$\tablenotemark{1}} &
   \colhead{B-GPA} &
   \colhead{$\langle$P-GPA$\rangle$} \\
   \colhead{} &
   \colhead{} &
   \colhead{(arcmin)} &
   \colhead{(deg)} &
   \colhead{(deg)} \\
   \colhead{(1)} &
   \colhead{(2)} &
   \colhead{(3)} &
   \colhead{(4)} &
   \colhead{(5)} \\
	}
	\input{latex_table3.txt}
	\tablenotetext{1}{Data taken from \citet{2006ApJ...649..759C}.}
	\tablenotetext{2}{Bubble fitting failed, GPA measured as described in text.}
\end{longtable}


\end{document}

%% file: hii_table2.txt
N11 		& 54.0 $\pm$ 0.1 	& 16.5 $\pm$ 0.1 	& 360 $\pm$ 60 	& 11.63 $\pm$ 1.16 	& 49 $\pm$ 13 	& 4.57 & 3.28 & O3 \\
N20 		& 39.1 $\pm$ 0.2 	& 19.5 $\pm$ 0.5 	& 130 $\pm$ 20 	& 12.72 $\pm$ 1.27 	& 21 $\pm$ 5.3 	& 4.20 & 3.16 & O6 \\
N23a 		& 42.6 $\pm$ 0.6 	& 23.2 $\pm$ 1.3 	& 390 $\pm$ 30 	& 12.51 $\pm$ 1.25 	& 61 $\pm$ 13 	& 1.42 & 1.31 & O3 \\
N23b 		& 61.9 $\pm$ 7.7 	& 35.6 $\pm$ 9.9 	& 390 $\pm$ 30 	& 11.49 $\pm$ 1.15 	& 51 $\pm$ 11 	& 1.30 & 1.20 & O3 \\
N25 		& 37.8 $\pm$ 0.1 	& 19.9 $\pm$ 0.2 	& 170 $\pm$ 20 	& 12.81 $\pm$ 1.28 	& 28 $\pm$ 6.5 	& 3.10 & 2.02 & O5 \\
N27a 		& 60.4 $\pm$ 0.1 	& 18.2 $\pm$ 0.3 	& 190 $\pm$ 40 	& 11.54 $\pm$ 1.15 	& 25 $\pm$ 7.4 	& 3.56 & 2.95 & O6 \\
N27b 		& 118 $\pm$ 0.3 	& 22.7 $\pm$ 0.8 	& 190 $\pm$ 40 	& 8.14 $\pm$ 0.81 	& 13 $\pm$ 3.7 	& 2.51 & 2.08 & O6 \\
N31a 		& 114.3 $\pm$ 0.3 	& 28.4 $\pm$ 0.8 	& 340 $\pm$ 80 	& 7.77 $\pm$ 0.78 	& 21 $\pm$ 6.3 	& 1.61 & 1.24 & O5 \\
N31b 		& 41.9 $\pm$ 0.3 	& 23.6 $\pm$ 0.8 	& 340 $\pm$ 80 	& 12.40 $\pm$ 1.24 	& 52 $\pm$ 16	& 2.56 & 1.98 & O3 \\
N42 		& 100.9 $\pm$ 0.1 	& 20.0 $\pm$ 0.2 	& 170 $\pm$ 10 	& 8.90 $\pm$ 0.89 	& 13 $\pm$ 2.8 	& 1.73 & 1.14 & O7 \\
&&&&&&&&  \\
N50 		& 67.7 $\pm$ 0.1 	& 21.7 $\pm$ 0.3 	& 450 $\pm$ 40 	& 10.46 $\pm$ 1.05 	& 49 $\pm$ 11 	& 4.84 & 4.29 & O4 \\
N53a		& 43.3 $\pm$ 0.1 	& 19.6 $\pm$ 0.1 	& 480 $\pm$ 50 	& 11.63 $\pm$ 1.16 	& 65 $\pm$ 15 	& 2.50 & 2.47 & O3 \\
N53b 		& 101.6 $\pm$ 0.8 	& 22.6 $\pm$ 1.9 	& 480 $\pm$ 50 	& 7.27 $\pm$ 0.73 	& 25 $\pm$ 5.8 	& 1.57 & 1.54 & O5 \\
N56 		& 77.4 $\pm$ 0.1 	& 21.0 $\pm$ 0.3 	& 270 $\pm$ 20 	& 9.29 $\pm$ 0.93 	& 23 $\pm$ 5.0 	& 2.72 & 2.67 & O6 \\
N57 		& 30.1 $\pm$ 0.5 	& 22.7 $\pm$ 1.1 	& 30 $\pm$ 10 	& 12.24 $\pm$ 1.22 	& 4.4 $\pm$ 1.7	& 1.10 & 0.89 & O8 \\
N60 		& 50 $\pm$ 0.2 		& 17.3 $\pm$ 0.4 	& 70 $\pm$ 10 	& 10.84 $\pm$ 1.08 	& 8.2 $\pm$ 2.0	& 2.02 & 1.51 & O7 \\
N62 		& 62.9 $\pm$ 0.1 	& 18.5 $\pm$ 0.3 	& 310 $\pm$ 40 	& 9.92 $\pm$ 0.99 	& 31 $\pm$ 7.3 	& 4.09 & 3.80 & O5 \\
N66a 		& 38.1 $\pm$ 1.0 	& 31.0 $\pm$ 2.0 	& 90 $\pm$ 20 	& 11.36 $\pm$ 1.14 	& 12 $\pm$ 3.5 	& 1.52 & 1.22 & O6 \\
N66b 		& 68.6 $\pm$ 1.7 	& 24.3 $\pm$ 4.2 	& 90 $\pm$ 20 	& 9.35 $\pm$ 0.94 	& 7.9 $\pm$ 2.4	& 1.25 & 1.01 & O7 \\
N66c 		& 92.0 $\pm$ 0.7 	& 19.1 $\pm$ 2.1 	& 90 $\pm$ 20 	& 6.94 $\pm$ 0.69 	& 4.3 $\pm$ 1.3	& 0.93 & 0.75 & B0 \\
&&&&&&&&  \\
N67 		& 57.5 $\pm$ 0.1 	& 21.2 $\pm$ 0.1 	& 470 $\pm$ 30 	& 10.06 $\pm$ 1.01 	& 48 $\pm$ 10 	& 2.20 & 2.02 & O4 \\
N73 		& 60.9 $\pm$ 0.2 	& 28.2 $\pm$ 0.4 	& 210 $\pm$ 20 	& 9.16 $\pm$ 0.92 	& 18 $\pm$ 4.0 	& 2.06 & 1.92 & O5 \\
N75 		& 42.1 $\pm$ 0.2 	& 21.4 $\pm$ 0.5 	& 40 $\pm$ 10 	& 10.46 $\pm$ 1.05 	& 4.4 $\pm$ 1.4	& 1.89 & 1.55 & O8 \\
N80 		& 20.7 $\pm$ 0.2 	& 30.7 $\pm$ 0.5 	& 620 $\pm$ 80 	& 11.18 $\pm$ 1.12 	& 77 $\pm$ 18 	& 6.22 & 4.52 & O3 \\
N90 		& 70.5 $\pm$ 0.1 	& 20.8 $\pm$ 0.2 	& 360 $\pm$ 60 	& 6.14 $\pm$ 0.61 	& 14 $\pm$ 3.5 	& 2.73 & 2.71 & O7 \\
N92 		& 62.5 $\pm$ 0.2 	& 17.0 $\pm$ 0.6 	& 140 $\pm$ 40 	& 7.29 $\pm$ 0.73 	& 7.4 $\pm$ 2.6	& 3.59 & 2.44 & O7 \\
N95 		& 52.5 $\pm$ 0.1 	& 20.6 $\pm$ 0.2 	& 610 $\pm$ 30 	& 8.12 $\pm$ 0.81 	& 40 $\pm$ 8.3 	& 4.97 & 3.96 & O4 \\
N96 		& -44.4 $\pm$ 0.1	& 29.3 $\pm$ 0.3 	& 450 $\pm$ 40 	& 14.65 $\pm$ 1.46 	& 97 $\pm$ 21 	& 1.88 & 1.55 & O3 \\
N98 		& 56.9 $\pm$ 0.1 	& 21.0 $\pm$ 0.2 	& 390 $\pm$ 40 	& 6.96 $\pm$ 0.70 	& 19 $\pm$ 4.2 	& 2.31 & 2.29 & O6 \\
N105 		& -1.1 $\pm$ 0.2 	& 18.3 $\pm$ 0.4 	& 200 $\pm$ 30 	& 10.62 $\pm$ 1.06 	& 23 $\pm$ 5.6 	& 2.83 & 1.86 & O6 \\
N110 		& 7.9 $\pm$ 0.4 	& 25.7 $\pm$ 0.9 	& 70 $\pm$ 30 	& 9.59 $\pm$ 0.96 	& 6.4 $\pm$ 3.0	& 1.60 & 1.48 & O8 \\
N115 		& 23.9 $\pm$ 0.3 	& 29.9 $\pm$ 0.7 	& 1250 $\pm$ 800& 8.14 $\pm$ 0.81 	& 83 $\pm$ 56 	& 7.58 & 5.97 & O3 \\
N122 		& 45.5 $\pm$ 0.3 	& 19.4 $\pm$ 0.6 	& 110 $\pm$ 30 	& 4.74 $\pm$ 0.47 	& 2.5 $\pm$ 0.8	& 0.73 & 0.51 & O9 \\

%% file: latex_table3.txt
&\multicolumn{3}{c}{Outside GPIPS Region}& \\
\hline
N1  &      0.47 &      0.93 & $ 54\pm   15$ & $\cdots$ \\
N2  &      0.56 &      6.96 & $108\pm   17$ & $\cdots$ \\
N3  &      0.87 &      0.99 & $ 32\pm    3$ & $\cdots$ \\
N4  &      0.53 &      1.95 & $ 64\pm   14$ & $\cdots$ \\
N5  &      0.79 &      3.71 & $ 39\pm    8$ & $\cdots$ \\
N6  &      0.64 &      5.94 & $ 17\pm   10$ & $\cdots$ \\
N7  &      0.60 &      0.47 & $135\pm    3$ & $\cdots$ \\
N8  &      0.00 &      0.17 & $ 58\pm  180$ & $\cdots$ \\
N9  &      0.71 &      0.59 & $133\pm    5$ & $\cdots$ \\
N10 &      0.80 &      1.41 & $137\pm    4$ & $\cdots$ \\
&&&&\\
N11 &      0.70 &      1.35 & $ 60\pm    4$ & $\cdots$ \\
N12 &      0.50 &      4.48 & $ 76\pm   14$ & $\cdots$ \\
N13 &      0.81 &      0.57 & $ 75\pm    4$ & $\cdots$ \\
N14 &      0.77 &      1.30 & $158\pm    5$ & $\cdots$ \\
N15 &      0.75 &      1.57 & $161\pm    4$ & $\cdots$ \\
N16 &      0.59 &      2.42 & $ 35\pm   14$ & $\cdots$ \\
N17 &      0.68 &      0.42 & $152\pm    1$ & $\cdots$ \\
N18 &      0.35 &      6.61 & $ 64\pm   44$ & $\cdots$ \\
N19 &      0.41 &      3.91 & $  4\pm   21$ & $\cdots$ \\
\hline
&\multicolumn{3}{c}{Inside GPIPS Region}& \\
\hline
N20 &      0.66 &      1.13 & $ 82\pm    6$ & $94.4\pm0.1$\\
N21 &      0.72 &      2.30 & $ 67\pm    8$ & $89.8\pm0.1$\\
N22 &      0.67 &      1.71 & $ 99\pm    8$ & $87.7\pm0.1$\\
N23 &      0.39 &      0.39 & $ 19\pm   15$ & $91.4\pm0.1$\\
N24 &      0.47 &      7.00 & $163\pm   23$ & $92.8\pm0.1$\\
N25 &      0.76 &      0.83 & $8.8\pm    3$ & $87.0\pm0.1$\\
N26 &      0.63 &      0.50 & $136\pm    3$ & $89.1\pm0.1$\\
N27 &      0.56 &      1.06 & $115\pm    9$ & $94.3\pm0.1$\\
N28 &      0.45 &      0.48 & $ 73\pm   17$ & $83.7\pm0.1$\\
N29 &      0.74 &      2.82 & $ 27\pm    8$ & $101.6\pm0.3$\\
N30\tablenotemark{2} &      0.57 &      0.99 & $ 55\pm   19$ & $96.5\pm0.2$\\
&&&&\\
N31 &      0.63 &      0.71 & $125\pm    6$ & $94.1\pm0.2$\\
N32 &      0.42 &      0.37 & $180\pm    5$ & $91.1\pm0.1$\\
N33 &      0.45 &      0.39 & $ 37\pm   10$ & $86.4\pm0.1$\\
N34 &      0.60 &      1.07 & $ 37\pm    8$ & $77.8\pm0.1$\\
N35 &      0.81 &      3.64 & $ 42\pm    5$ & $90.4\pm0.1$\\
N36 &      0.75 &      2.93 & $ 23\pm    6$ & $85.9\pm0.2$\\
N37 &      0.84 &      2.07 & $ 16\pm    4$ & $83.3\pm0.1$\\
N38 &      0.52 &      0.58 & $167\pm   15$ & $80.4\pm0.1$\\
N39 &      0.68 &      1.95 & $ 75\pm    9$ & $82.0\pm0.2$\\
&&&&\\
N40 &      0.70 &      1.23 & $ 71\pm    5$ & $83.4\pm0.1$\\
N41 &      0.25 &      0.42 & $ 18\pm   70$ & $80.4\pm0.1$\\
N42 &      0.76 &      0.67 & $150\pm    6$ & $83.9\pm0.1$\\
N43 &      0.58 &      0.61 & $1.3\pm    9$ & $89.1\pm0.1$\\
N44 &      0.55 &      1.10 & $165\pm    9$ & $91.4\pm0.1$\\
N45\tablenotemark{2} &      0.78 &      1.53 & $ 50\pm   15$ & $90.0\pm0.2$\\
N46 &      0.38 &      1.19 & $9.8\pm   21$ & $69.6\pm0.2$\\
N47 &      0.73 &      2.40 & $ 54\pm    6$ & $100.2\pm0.2$\\
N48 &      0.35 &      0.87 & $142\pm   18$ & $110.2\pm0.2$\\
N49 &      0.63 &      1.32 & $ 90\pm    9$ & $104.9\pm0.3$\\
&&&&\\
N50 &      0.46 &      1.59 & $ 24\pm   17$ & $116.4\pm0.3$\\
N51 &      0.25 &      1.78 & $ 46\pm   64$ & $91.8\pm0.4$\\
N52 &      0.79 &      2.22 & $ 48\pm    5$ & $76.5\pm0.2$\\
N53 &      0.15 &      0.74 & $134\pm  133$ & $83.7\pm0.2$\\
N54\tablenotemark{2} &      0.70 &      1.88 & $ 80\pm   16$ & $67.4\pm0.2$\\
N55 &      0.67 &      0.77 & $113\pm    6$ & $99.2\pm0.4$\\
N56 &      0.19 &      0.99 & $ 74\pm  102$ & $90.1\pm0.2$\\
N57 &      0.57 &      0.31 & $ 75\pm    7$ & $84.3\pm0.2$\\
N58 &      0.62 &      0.17 & $178\pm    1$ & $90.6\pm0.2$\\
N59 &      0.68 &      7.03 & $ 23\pm    9$ & $87.1\pm0.1$\\
&&&&\\
N60 &      0.66 &      0.64 & $155\pm    4$ & $70.3\pm0.3$\\
N61 &      0.30 &      3.28 & $128\pm   62$ & $79.2\pm0.3$\\
N62 &      0.38 &      1.41 & $ 36\pm   26$ & $75.6\pm0.2$\\
N63\tablenotemark{2} &      0.68 &     11.89 & $ 25\pm   16$ & $90.5\pm0.1$\\
N64 &      0.69 &      5.08 & $ 42\pm    9$ & $93.1\pm0.2$\\
N65 &      0.49 &      2.02 & $ 10\pm   16$ & $81.2\pm0.2$\\
N66 &      0.62 &      0.46 & $ 90\pm    7$ & $94.3\pm0.2$\\
N67 &      0.38 &      0.75 & $ 95\pm   13$ & $103.0\pm0.1$\\
N68 &      0.72 &      5.17 & $9.0\pm   10$ & $94.2\pm0.1$\\
N69 &      0.00 &      8.42 & $6.3\pm  180$ & $93.5\pm0.1$\\
&&&&\\
N70 &      0.46 &      0.50 & $ 38\pm   17$ & $\cdots$ \\
N71 &      0.53 &      7.08 & $ 74\pm   16$ & $95.5\pm0.1$\\
N72 &      0.36 &      0.92 & $ 41\pm   26$ & $92.0\pm0.2$\\
N73 &      0.34 &      0.77 & $ 41\pm   24$ & $84.1\pm0.2$\\
N74 &      0.59 &      1.27 & $ 52\pm    7$ & $90.7\pm0.3$\\
N75 &      0.57 &      0.62 & $172\pm    6$ & $88.9\pm0.3$\\
N76 &      0.65 &      3.73 & $ 77\pm   10$ & $96.1\pm0.2$\\
N77 &      0.33 &      1.16 & $ 74\pm   31$ & $68.7\pm0.1$\\
N78\tablenotemark{2} &      0.68 &      0.31 & $135\pm   16$ & $71.1\pm0.1$\\
N79 &      0.64 &      1.27 & $169\pm    9$ & $65.4\pm0.2$\\
&&&&\\
N80 &      0.69 &      1.90 & $ 62\pm    9$ & $72.1\pm0.2$\\
N81 &      0.70 &      9.24 & $ 73\pm    7$ & $64.8\pm0.1$\\
N82 &      0.13 &      1.47 & $119\pm  180$ & $77.3\pm0.3$\\
N83\tablenotemark{2} &      0.65 &      0.31 & $155\pm   17$ & $73.1\pm0.2$\\
N84 &      0.69 &      1.19 & $3.1\pm    6$ & $77.0\pm0.3$\\
N85 &      0.57 &      0.46 & $104\pm    3$ & $85.3\pm0.3$\\
N86 &      0.46 &      0.26 & $ 50\pm    2$ & $85.2\pm0.3$\\
N87 &      0.61 &      0.31 & $ 32\pm    6$ & $87.6\pm0.3$\\
N88 &      0.30 &      1.37 & $176\pm   32$ & $94.7\pm0.4$\\
N89 &      0.47 &      0.92 & $154\pm   15$ & $\cdots$ \\
&&&&\\
N90 &      0.13 &      1.53 & $122\pm  180$ & $\cdots$ \\
N91 &      0.74 &      5.19 & $ 40\pm    8$ & $107.1\pm0.3$\\
N92 &      0.73 &      2.03 & $ 17\pm    7$ & $\cdots$ \\
N93 &      0.68 &      0.73 & $  9\pm    8$ & $\cdots$ \\
N94 &      0.74 &      3.85 & $ 96\pm    6$ & $85.3\pm0.2$\\
N95 &      0.60 &      2.13 & $125\pm    9$ & $87.2\pm0.5$\\
N96 &      0.59 &      0.40 & $ 11\pm    3$ & $97.0\pm0.2$\\
N97 &      0.67 &      4.23 & $ 95\pm    8$ & $86.0\pm0.6$\\
N98 &      0.11 &      1.37 & $ 43\pm  180$ & $97.2\pm0.3$\\
N99 &      0.64 &      4.20 & $177\pm   11$ & $95.4\pm0.3$\\
&&&&\\
N100 &      0.63 &      4.85 & $ 54\pm   11$ & $82.1\pm0.2$\\
N101\tablenotemark{2} &      0.52 &      0.99 & $125\pm   21$ & $81.9\pm0.2$\\
N102 &      0.12 &      1.82 & $171\pm  180$ & $82.2\pm0.3$\\
N103 &      0.63 &      0.63 & $ 91\pm    9$ & $82.0\pm0.3$\\
N104 &      0.64 &      0.77 & $ 34\pm    8$ & $81.0\pm0.2$\\
N105 &      0.75 &      0.87 & $103\pm    5$ & $79.8\pm0.2$\\
N106 &      0.65 &      0.39 & $ 36\pm   10$ & $84.8\pm0.2$\\
N107 &      0.62 &     11.59 & $110\pm   12$ & $81.1\pm0.1$\\
N108 &      0.84 &      4.39 & $137\pm    4$ & $77.0\pm0.2$\\
N109 &      0.61 &     14.57 & $109\pm   12$ & $78.7\pm0.1$\\
&&&&\\
N110 &      0.38 &      0.55 & $122\pm   19$ & $71.6\pm0.1$\\
N111 &      0.59 &      1.00 & $ 44\pm    8$ & $71.4\pm0.1$\\
N112 &      0.74 &      0.28 & $143\pm    7$ & $71.6\pm0.1$\\
N113 &      0.80 &      0.57 & $ 38\pm    3$ & $72.4\pm0.1$\\
N114 &      0.20 &      1.35 & $ 72\pm   75$ & $71.9\pm0.1$\\
N115 &      0.62 &      3.20 & $ 80\pm   12$ & $81.3\pm0.2$\\
N116 &      0.61 &      0.79 & $118\pm    5$ & $77.8\pm0.2$\\
N117 &      0.14 &      1.41 & $ 48\pm  180$ & $77.4\pm0.2$\\
N118 &      0.58 &      0.32 & $122\pm    6$ & $74.1\pm0.2$\\
N119 &      0.48 &      4.31 & $178\pm   20$ & $77.6\pm0.2$\\
&&&&\\
N120 &      0.33 &      1.26 & $ 99\pm   32$ & $72.7\pm0.3$\\
N121 &      0.55 &      0.49 & $144\pm    9$ & $70.2\pm0.4$\\
N122 &      0.72 &      0.53 & $177\pm    3$ & $79.1\pm0.4$\\
\hline
&\multicolumn{3}{c}{Outside GPIPS Region}& \\
\hline
N123 &      0.73 &      1.47 & $106\pm    6$ & $\cdots$ \\
N124 &      0.35 &      1.60 & $161\pm   33$ & $\cdots$ \\
N125 &      0.44 &      0.74 & $142\pm   19$ & $\cdots$ \\
N126 &      0.61 &      1.82 & $ 69\pm    9$ & $\cdots$ \\
N127 &      0.20 &      2.94 & $ 24\pm  178$ & $\cdots$ \\
N128 &      0.70 &      3.38 & $ 60\pm    8$ & $\cdots$ \\
N129 &      0.70 &      3.27 & $ 13\pm   17$ & $\cdots$ \\
&&&&\\
N130 &      0.47 &      0.98 & $169\pm   13$ & $\cdots$ \\
N131 &      0.47 &      6.18 & $ 77\pm   21$ & $\cdots$ \\
N132 &      0.36 &      0.21 & $ 86\pm    7$ & $\cdots$ \\
N133 &      0.78 &      1.83 & $ 73\pm    5$ & $\cdots$ \\
N134 &      0.83 &      0.58 & $ 79\pm    3$ & $\cdots$ \\